\documentclass{kluwer}    % Specifies the document style.

\usepackage{psfig}

\def\etal{{et\,al.}}
\def\asec{\ifmmode ^{\prime\prime}\else$^{\prime\prime}$\fi}
\def\farcs{\hbox{$.\!\!^{\prime\prime}$}}  % Fractions of arcseconds
\def\msun{$M_{\odot}$}
\newbox\grsign \setbox\grsign=\hbox{$>$}
\newdimen\grdimen \grdimen=\ht\grsign
\newbox\laxbox \newbox\gaxbox
\setbox\gaxbox=\hbox{\raise.5ex\hbox{$>$}\llap
     {\lower.5ex\hbox{$\sim$}}}\ht1=\grdimen\dp1=0pt
\setbox\laxbox=\hbox{\raise.5ex\hbox{$<$}\llap
     {\lower.5ex\hbox{$\sim$}}}\ht2=\grdimen\dp2=0pt

\def\lax{$\mathrel{\copy\laxbox}$}

\begin{document}                                                                                   
\begin{article}
\begin{opening}         
\title{Identification of the donor in GRS 1915+105
%\thanks{Footnote 
%            to the title with the `thanks' command.}
} 
\author{J. \surname{Greiner}} 
\author{M.J. \surname{McCaughrean}}
\runningauthor{Greiner et al.}
\runningtitle{Donor of GRS 1915+105}
\institute{Astrophysical Institute Potsdam, 14482 Potsdam, Germany}
\author{J.G. \surname{Cuby}}
\institute{ESO, Alonso de C\'ordova 3107, Santiago 19, Chile}
\author{A.J. \surname{Castro-Tirado}}
\institute{Instituto de Astrof\'{\i}sica de Andaluc\'{\i}a,
                             E-18080 Granada, Spain}
\author{R.E. \surname{Mennickent}}
\institute{Universidad de Concepci\'on, Casilla 160-C Concepcion, Chile}

%\date{April 15, 1993}

\begin{abstract}
We report on the results of medium-resolution spectroscopy of GRS 1915+105
in the H and K band using the 8m VLT at ESO.
We clearly identify absorption bandheads from $^{12}$CO and $^{13}$CO.
Together with other features this results in a classification of the 
donor as a K-M\,III star.
\end{abstract}

\end{opening}           

\section{Introduction}  

GRS 1915+105 (Castro-Tirado \etal\ 1994) is the prototypical microquasar, 
a galactic X-ray binary
ejecting plasma clouds at v$\approx$0.92\,c (Mirabel \& Rodriguez 1994).
It exhibits unique X-ray variability patterns (Greiner \etal\ 1996) 
which have been interpreted as
instabilities leading to an infall of parts of the
inner accretion disk (Belloni \etal\ 1997).
It is therefore of great importance to know some details about the
system components in order to understand the conditions which lead to
these unique characteristics.
%  -- a galactic source at distance d $\approx$ 9--12 kpc 
%      (Fender \etal\ 1999)\\
%  -- at bII = -0\fdg2 behind a column of A$_V$ $\approx$ 27 mag \\
%Some correlated X-ray/infrared/radio variability patterns have also been 
%related to jet formation  (Mirabel \etal\ 1998).
%Three different sources of infrared (IR) radiation are expected from a 
%X-ray binary,
%namely thermal emission from the outer part of accretion disk, thermal 
%emission from the companion star or synchrotron emission from the jet(s).
%Infrared variability could be caused by a variety of processes, among them
%changes of the mass transfer rate from the secondary to the accretion disk,
%a changing aspect of the illuminated secondary,
%a changing illumination of the disk and/or secondary, 
%sporadic jet ejection and infrared synchrotron emission, or
%free-free emission from an X-ray driven wind.
%Previous IR observations of GRS 1915+105 include, besides
%a few single measurements since 1993, 
%some few hrs photometry in conjunction with X-ray/radio measurements 
%     showing 3 different types of IR-behaviour 
%(Eikenberry \etal\ 2000), and
%a 2 month monitoring in 1996 (Bandyopadhyay \etal\ 1998).

\section{Observations and Results}

We obtained H and K band infrared spectroscopy of GRS 1915+105
with the aim of searching for absorption signatures due to the donor. Since
GRS 1915+105 is a strongly variable infrared source, believed to 
predominantly caused by synchrotron emission of ejected material
(Eikenberry \etal\ 2000, Greiner \etal\ 2001), this required high 
signal-to-noise as well as good spectral resolution in order to beat the 
strong veiling. We therefore used the infrared spectrometer ISAAC 
on the 8m VLT Antu telescope on Paranal (ESO, Chile). 

 \begin{figure}[th]
   \centering{
   \hspace{-0.01cm}
%   \vbox{\psfig{figure=grs1915_K1spec.ps,width=0.73\textwidth,%
%         bbllx=2.1cm,bblly=2.6cm,bburx=17.2cm,bbury=12.1cm,clip=}}\par
%   \hspace{-0.01cm}
%   \vbox{\psfig{figure=grs1915_K2spec.ps,width=0.73\textwidth,%
%         bbllx=2.1cm,bblly=2.6cm,bburx=17.2cm,bbury=12.1cm,clip=}}\par
%   \hspace{-0.01cm}
%   \vbox{\psfig{figure=grs1915_K3spec.ps,width=0.73\textwidth,%
%         bbllx=2.4cm,bblly=5.2cm,bburx=14.5cm,bbury=12.1cm,clip=}}\par
   \vbox{\psfig{figure=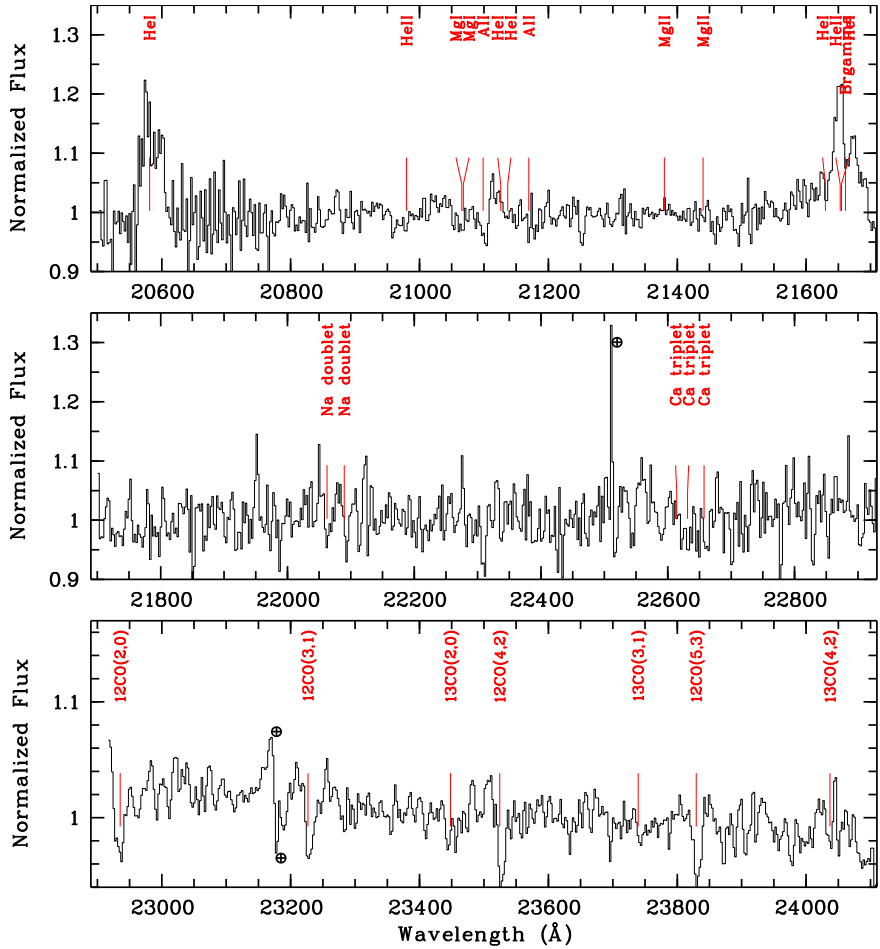,width=0.99\textwidth,%
         bbllx=2.2cm,bblly=7.7cm,bburx=19.2cm,bbury=26.2cm,clip=}}\par
   \vspace{-0.2cm}
   \caption[spec]{Spectrum of GRS 1915+105 in the K band. The top two panels
          correspond to 3000 sec exposure, while the bottom panel shows 
          a sum of 4 images with 2000 sec exposure each. All spectra were 
          rebinned by a factor of 2.
          \vspace{-0.15cm}
          }}
         \label{spec}
\end{figure}

The short wavelength (0.9--2.5 $\mu$m) arm of ISAAC is
equipped with a 1024$\times$1024 pixel Rockwell HgCdTe array
with an image scale of 0\farcs147/pixel.
Using the grating with the highest possible spectral resolution ($\sim$3000
for a 1\asec\ slit) yields
0.8\AA/pixel in the H band and 1.2\AA/pixel in the K band.
Observations were performed in 1 (2) adjacent H bands, and 2 (3) 
adjacent K bands on 20/21 July 1999 (24/25 July 2000).

Science exposures consisted of several 250--300 sec individual exposures
which were dithered along the slit by $\pm$30\asec.
In order to correct for atmospheric absorption, the nearby star HD 179913
(A0\,V) was observed either before or after each science exposure.
The initial data reduction steps like debiasing, flatfielding and co-adding
were performed within the {\em Eclipse} package (Devillard 2000).
The extraction and wavelength calibration was done using an optimal extraction
routine within the MIDAS package.

The spectra of the 3 grating settings covering the K band are shown in Fig. 1.
The top panel shows the strong emission lines of Br\,$\gamma$ and He\,I
known already from previous low-resolution spectroscopy
(Castro-Tirado \etal\ 1996). In addition, we find
for the first time several absorption lines which allow us to roughly
identify the donor in GRS 1915+105.
The lower panel clearly shows $^{12}$CO absorption band heads characteristic 
of a low temperature ($T< 7000$ K) star (e.g. Kleinmann \& Hall 1986). 
Though weak, we also identify the 
$^{13}$CO (2,0) and $^{13}$CO (3,1) transitions, indicating a luminosity
class III or brighter (e.g. Wallace \& Hinkle 1997). 
We also identify the Na doublet (2.20624/2.20897 $\mu$m), and possibly
the Ca triplet (2.26141/2.26311/2.26573\,$\mu$m), Al\,I (2.10988 $\mu$m)
and the Mg\,I doublet (2.10655/2.10680\,$\mu$m) in absorption.
Note that the CN doublet (2.0910/2.0960\,$\mu$m), which in supergiants
is more prominent than Al/Mg, is not detected. Thus, we conclude that
the donor in GRS 1915+105 is a late-type K-M giant.

We have tried to confirm the luminosity class more quantitatively
by using the veiling-independent indicator \\
 $~~~~~r = \log [EW(^{12}{\rm CO} (2,0))/(EW({\rm Na}) + EW({\rm Ca}))]$ \\
(Ram\'irez \etal\ 1997).
Because of the low significance of the Ca triplet our measurement
has a large error: $r = 0.25 \pm 0.20$. This value falls in between
the ranges covered by dwarfs (--0.2\lax $r$ \lax 0.0) and
giants (0.4\lax $r$ \lax 0.6) (Ram\'irez \etal\ 1997).
The ratio of equivalent widths of $^{12}$CO to $^{13}$CO 
which depends on luminosity class (Campbell \etal\ 1990), has been measured
for the seven transitions covered (lower panel of Fig. 1)
to $\sim 3 \pm 1$, again supporting a giant classification.

Unfortunately, the H band spectra are of too low S/N ratio to detect OH or CO,
thus not allowing us to use the veiling-independent temperature/luminosity
discriminants as proposed by Meyer \etal\ (1998). However, the veiling was 
roughly determined by comparison of our flux-calibrated 2.35\,$\mu$m spectrum 
of GRS 1915+105 with that of a K2\,III standard star, observed with the same
settings. We adopted a 13 mag flat continuum and added the K2\,III star
spectrum scaled to a brightness in the range 13.5--17.5 mag (Fig. 2). 
A comparison
with the GRS 1915+105 spectrum gives a (not extinction-corrected) magnitude of 
K = 14.5--15.0 mag for the donor.
With a distance of $\sim$11 kpc and a $A_{\rm K} = 3$ mag extinction correction
this implies an absolute magnitude of 
$M_{\rm K}$ = --2...--3 mag, consistent with the giant classification.

 \begin{figure}[th]
   \centering{
   \hspace{-0.01cm}
   \vbox{\psfig{figure=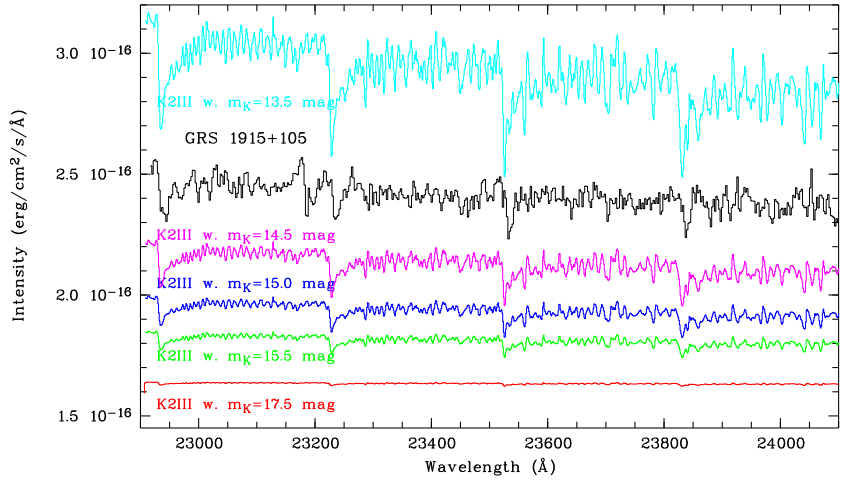,width=0.95\textwidth,%
         bbllx=0.9cm,bblly=2.6cm,bburx=17.2cm,bbury=12.1cm,clip=}}\par
   \vspace{-0.15cm}
   \caption[veil]{Spectrum of GRS 1915+105 in the 2.3--2.4 $\mu$m range (second
          from top, not extinction-corrected), compared to the spectrum
          of the K2III star HD 202135 scaled to different magnitudes (as
          marked on each spectrum) which is veiled by a 13 mag flat continuum. 
          This implies that the donor of GRS 1915+105 has a
          magnitude of 14.5--15 mag prior to extinction correction.
          }}
         \label{veil}
 \end{figure}

%These findings do not contradict the results of 
%\citeauthor{Chin88-book} (on which \shortcite{Bunt} based their 
%production kinetics) and of \citeyear{ChinThesis} which were obtained for

%\section{Results and Discussion}

This identification of the donor of GRS 1915+105 as a K-M giant
implies a rather narrow mass range of 1.0--1.5 \msun. 
Typical spherical mass-loss rates of such stars are much too low to sustain
the high accretion luminosity of GRS 1915+105 via accretion from 
the donor's stellar wind. We therefore suggest (not too surprisingly)
that accretion should occur via Roche lobe overflow.
We finally note that while our identification
contradicts the findings of Mart\'i \etal\ (2000) who argue for a massive
OB-type companion, it is consistent
with the constraints derived by Eikenberry \etal\ (2001).

The presence of clear donor absorption features will now allow 
a period search as well as a determination of the mass of the compact
object using radial velocity measurements of the donor.

%\acknowledgements
%And this is an acknowledgement

\end{article}
\end{document}